# Experimental Validation of Skull Acoustic Modelling Strategies for Transcranial Focused Ultrasound Simulation: A Cross-Comparison Study

Han Li[1,4], Aidan J Horner[2,3], Xinyu Zhang[5], Zhihong Huang[1]

[1] School of Physics, Engineering and Technology, University of York, York, YO10 5DD, United Kingdom

[2] Department of Psychology, University of York, York, YO10 5DD, United Kingdom

[3] York Biomedical Research Institute, University of York, York, YO10 5DD, United Kingdom

[4] School of Science and Engineering, University of Dundee, Dundee, DD1 4HN, United Kingdom

[5] School of Medicine, University of Dundee, Dundee, DD1 9SY, United Kingdom

## Abstract

### Objective

Accurate acoustic modelling of the skull is pivotal for simulation-guided transcranial focused ultrasound (tFUS) applications. However, existing skull parameterization strategies vary widely in complexity and reported accuracy. This study systematically evaluates commonly used modelling approaches to determine their predictive performance across multiple frequencies and skull types.

### Approach

Five skull modelling strategies were assessed: two voxel-wise linear mapping models, one with fixed attenuation coefficients and one with Hounsfield unit (HU)-dependent attenuation coefficient; two three-layer models with fixed and skull-density ratio (SDR) dependent attenuation coefficient; and one single-layer fixed-parameter model. Nineteen regions of interest from five historical and two Thiel-embalmed human skulls were examined at 220 kHz, 680 kHz and 1000 kHz, bowl-surface source fields were reconstructed using holography, and intracranial pressure distributions were captured with a needle hydrophone. Measurements were compared with k-Wave simulations.

### Results

Average simulated peak pressure errors ($\varepsilon_P$) ranged from 20% to 31% across frequencies, whereas intensity errors ($\varepsilon_I$) reached 41% to 77%. Errors in -6 dB focal volume ranged from 11% to 67%, and focal position discrepancies were typically several millimeters. Simulation generally predicted smaller insertion losses than measured, indicating a tendency to underestimate skull-related attenuation and overestimate transmitted intracranial exposure. Thiel specimens produced larger prediction errors than historical bone ($\varepsilon_P$: 29% vs. 23%; $\varepsilon_I$: 80% vs. 52%). Among the five modelling strategies, the linear mapping method with fixed attenuation coefficient yielded the lowest frequency-averaged pressure error ($\varepsilon_P$: 23%), whereas this advantage did not extend consistently to other comparison metrics.

**Significance**

These results indicate that current skull modelling strategies may reproduce the gross intracranial beam pattern while still misestimating the delivered acoustic exposure, focal coverage and peak location. Such errors are directly relevant to tFUS dose selection, target coverage and safety-margin interpretation, underscoring the need for refined parameterization methods for reliable tFUS treatment planning.

# Introduction

Transcranial focused ultrasound (tFUS) stimulation for neuromodulation has attracted increasing research interest as a means of evoking reproducible physiological effects in targeted brain regions [1–3]. In humans, tFUS has been investigated across cortical and subcortical targets, with reported effects on sensory, cognitive, affective and network-level brain activity [4–7], supporting its potential as a spatially selective neuromodulation tool. In contrast to deep brain stimulation therapies, which deliver electrical impulses via implanted electrodes, ultrasound modulates neuronal responses through mechanical interactions, including the perturbation of the permittivity of mechano-sensitive ion channels [8]. Its non-invasive nature positions tFUS as a potential alternative to invasive neuromodulation techniques.

Relative to other non-invasive modalities, such as Transcranial Magnetic Stimulation (TMS) which is typically constrained to superficial cortical targets and offers limited spatial specificity [9], ultrasound enables tight spatial focusing and deeper penetration. Focusing can be achieved using a concave piezoceramic transducer, helmet shaped multi-element transducers [10], or phased arrays that impose controlled phase delays across flat elements [11]. The acoustic high-pressure zone is generally an elongated, cigar shaped volume whose exact dimensions depend on the geometrical shape and operating frequency of the transducer. Human tFUS neuromodulation studies are typically performed in the sub-MHz range, often at carrier frequencies of a few hundred kilohertz. In brain tissue, attenuation in this range is modest (on the order of approximately 1% to 7% per centimeter) [12], allowing sufficient and relatively predictable energy delivery to subcortical structures.

However, a principal obstacle to consistent transcranial energy delivery is the skull. Significant acoustic impedance mismatches at the scalp-outer table and inner table-dura mater interfaces cause substantial reflections, impeding transmission into the cranial cavity and trapping energy in the skull layers. The skull's heterogeneous highly porous architecture (two cortical tables separated by trabecular diploë) further induces scattering and absorption. Furthermore, typical tFUS sonication protocols deliver trains of tone bursts, with each burst often containing tens to hundreds of sinusoidal cycles [13,14]. When combined with multiple reflected paths in bone, these waveforms can produce complex intracranial reverberation and superposition phenomena [15]. The resulting constructive and destructive interference leads to uncertainty in ultrasound insertion loss, particularly at lower frequencies, and raises concerns regarding localized skull heating [16] if not appropriately managed.

In order to anticipate transcranial wave propagation and improve reproducibility of physiological effect across subjects, many research groups [17–21] have employed numerical acoustic simulations. Among these, the k-Wave Matlab toolbox [22] is widely used for solving pseudo-spectral time domain wave equations with flexible definition of custom acoustic sources and media. Achieving realistic predictions depends critically on accurate representation of both the acoustic source and the propagation medium. In free-field, the source can be characterized experimentally and validated directly against hydrophone measurements, for example using acoustic holography to reconstruct the radiating pressure field [23]. Once the skull is introduced, however, the dominant uncertainty shifts from the source to the skull representation. Skull properties are usually inferred through empirical transformations and remain comparatively under-validated, with multiple approaches in current use and no consensus on their relative accuracy.

Existing skull modelling strategies span a spectrum of complexity. At one end, homogeneous models represent the skull as a single uniform layer with globally assigned density, sound speed and attenuation coefficients [24,25]. Such models are attractive for analytical insights and rapid computation but neglect spatial heterogeneity and layer-specific properties. Layer structured models explicitly represent the inner and outer cortical tables and intervening trabecular layer, enabling assignment of layer-wise physical and acoustic parameters and partially capturing structural anisotropy [26,27]. CT-derived models seek higher realism by mapping computed tomography (CT) information (e.g., Hounsfield units, HU) to spatially varying acoustic parameters [19,28]. In these approaches, density and sound speed are linearly estimated via empirically determined relationships, while the attenuation coefficient may be inferred using local porosity metrics [13] or skull density ratio values (SDR) [29–31], with frequency-dependent power-law absorption. Each strategy represents a trade-off between fidelity, computational cost, and dependence on imaging quality.

Despite the widespread adoption of these techniques, their predictive performance has not been fully cross validated against one another and against experimental measurement across anatomically varied skull regions. This gap affects both dose estimation and localization: inaccurate pressure or intensity predictions may lead to incorrect driving-level selection, whereas inaccurate focal volume or peak-position predictions may affect target coverage and unintended exposure of neighbouring tissue. Here, we implemented five skull modelling strategies and assessed their ability to predict intracranial acoustic fields measured in vitro. The evaluation comprised 19 ROIs from five historical skulls and two Thiel-embalmed skulls, tested at 220 kHz, 680 kHz and 1000 kHz. This study therefore provides a systematic experimental validation of commonly used skull acoustic modelling strategies for tFUS simulation.

# Methods

## Skull selection and imaging

Ex-vivo skull specimens were sourced from the Centre for Anatomy and Human Identification, University of Dundee. To sample a range of morphologies and storage conditions, we selected three historical calvaria (S1-S3), two Theil-embalmed calvaria (S4-S5; with scalp and dura mater attached) and two historical hemisected skulls (S6-S7). Relative to in-vivo bone, historical material has reduced moisture and marrow/lipid content within the bone matrix, rehydration was therefore required to approximate fresh conditions. Although the Thiel embalmed calvaria were freshly excised, air ingress through cut edges can occur during handling and transportation. Accordingly, all specimens were soaked in distilled water and vacuum degassed overnight prior to experimentation.

Rehydrated skulls were scanned on a clinical CT (GE revolution EVO, US) using a head protocol filter. Volumes were reconstructed with in-plane pixel spacing of 0.44 mm and slice spacing of 0.62 mm with an interval of 0.31 mm. Prior to imaging, regions of interest (ROIs) were defined as follows: for each calvaria, three ROIs were placed over frontal, left parietal and right parietal bones; for each half-skull, two ROIs were placed over the parietal and temporal bones. ROI locations were chosen at random, independent of cranial morphology or prospective brain targets; consequently, some unfavorable sites (e.g., over the frontal crest and squamous sutures) were retained. In total, 19 ROIs were included across the seven specimens (see Figure 1).

To facilitate precise co-registration between imaging and subsequent acoustic measurements, a metallic ball bearing marker (diameter = 3 mm) was affixed at the center of ROI prior to scanning, which helped unambiguous identification of target location within the CT volume. An in-house developed skull analysis algorithm was used for global inspection of each skull, generating maps and summary statistics of skull thickness and the SDR for visual inspection.

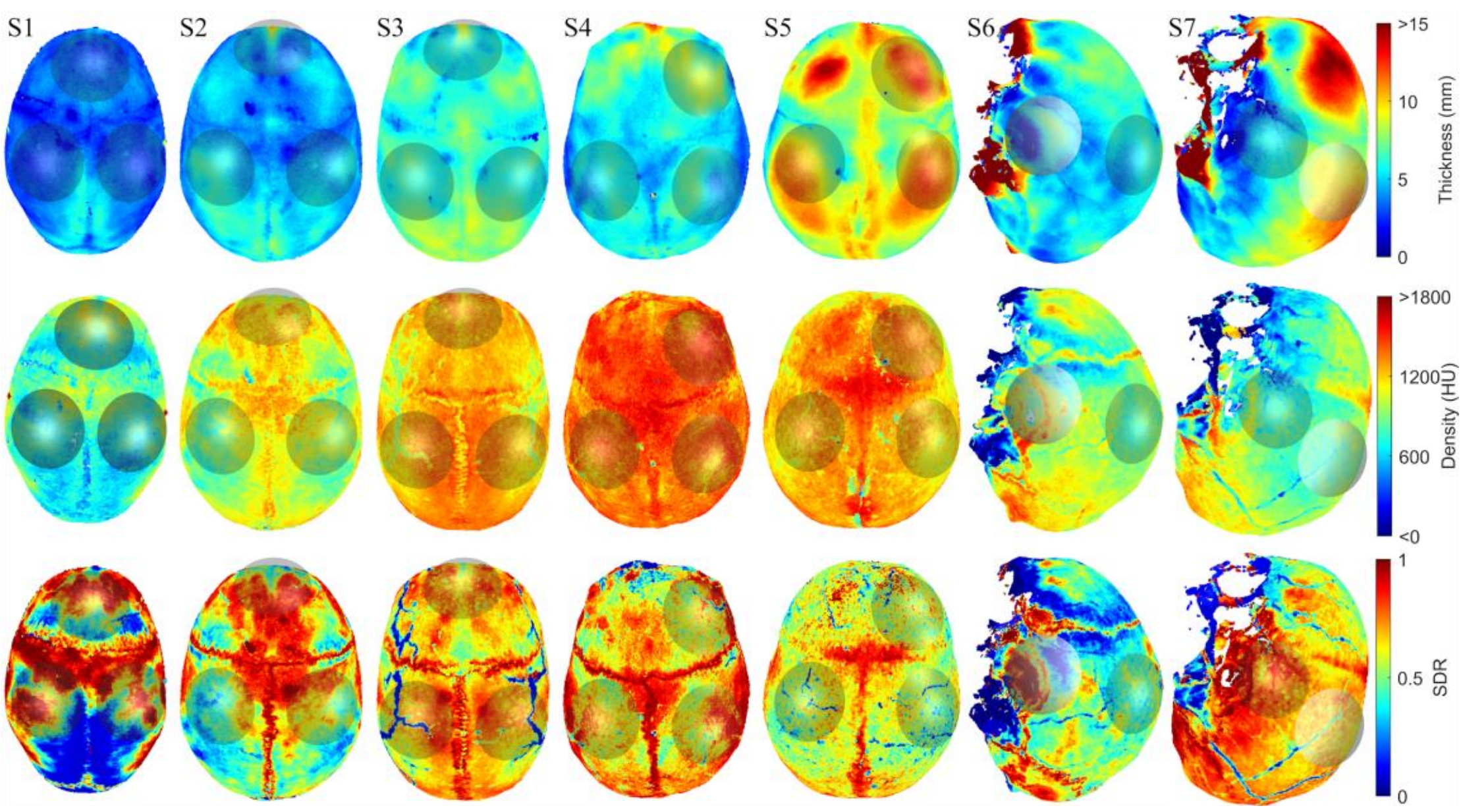

*Figure 1. Skull property maps, Upper row: Skull thickness maps of S1-S7. Middle row: the corresponding skull average HU density maps of S1-S7. Lower row: the corresponding SDR maps of S1-S7. Grey shades represent the ROI and the allocation of the acoustic source.*

## Skull acoustic modelling setup

Skull CT volumes were processed in MATLAB. For each ROI, a sub-volume with 60 mm×60 mm surface area was trimmed and placed orthogonally. The voxels were treated as bones where the HU was above +500 HU at air to the inner and outer bone table transitioning boundary. We neglected the presence of soft tissue attached on the Thiel samples to only model the bone condition. The five skull acoustic models were defined by the following transformation functions, with notation and units of $\rho$: mass density (kg/m$^3$), $v$: sound speed (m/s), and $a$: attenuation coefficient (dB/(cm·MHz$^2$)).

**1.** Linear mapping from the CT to density and sound speed with fixed attenuation coefficients (LM-Fa) [32]:

$$\rho_{skull} = 0.66HU_{skull} + 1011, \qquad HU_{skull} > 260$$
$$v_{skull} = 1.333\rho_{skull} + 166.7$$

$$a_{skull} = 13.3$$

**2.** Linear mapping from the CT to density and sound speed with dynamic attenuation coefficients (LM-Da) [13]:

$$\rho_{skull} = \frac{(\rho_{skull_max} - \rho_{water})HU_{skull}}{HU_{skull_max}} + \rho_{water}$$

$$v_{skull} = \frac{(v_{skull_max} - v_{water})(\rho_{skull} - \rho_{water})}{\rho_{skull_max} - \rho_{water}} + v_{water}$$

$$a_{skull} = \left(a_{skull_max} - a_{skull_min}\right) * \left(1 - \frac{HU_{skull} - HU_{skull_min}}{HU_{skull_max} - HU_{skull_min}}\right)^{\frac{1}{2}} + a_{skull_min}$$

Where $\rho_{skull_max} = 1900$, $\rho_{water} = 1000$, $v_{skull_max} = 3100$, $v_{water} = 1500\ (all\ simulations)$, $a_{skull_max} = 18.1$, $a_{skull_min} = 1.9.$

**3.** Outer cortical table, trabecular diploë, and inner cortical table segmented from CT; each layer assigned fixed properties (3L-F):

$$\rho_{skull_{cortical}} = 1850,\ \rho_{skull_trabecular} = 1700\ [25]$$

$$v_{skull_{cortical}} = 2800,\ v_{skull_trabecular} = 2300\ [25]$$

$$a_{skull_cortical} = 4, \qquad a_{skull_trabecular} = 16$$

**4.** Layered segmentation as 3L-F, density and sound speed were mapped from HU with local specificity; attenuation fixed for cortical bone and SDR-dependent for trabecular bone (3L-D) [29]:

$$\rho_{skull} = 0.66HU_{skull} + 1011, \qquad \rho_{CT} > 260$$

$$v_{skull} = 0.70HU_{skull} + 1730$$

$$a_{skull_cortical} = 4, \qquad a_{skull_trabecular} = 4 + 17(1 - SDR)$$

**5.** Single-layer skull represented as a homogeneous medium [24]:

$$\rho_{skull} = 1860$$

$$v_{skull} = 2890$$

$$a_{skull} = 7.3$$

## Acoustic source selection and characterization

Three single-element, concave focused ultrasound transducers (Precision Acoustics, UK, aperture diameter = 60 mm, radius of curvature (ROC) = 75 mm) were used in this study, each were operated near its center frequency 220 kHz, 680 kHz and 1000 kHz, covering the common bandwidth of interest in neuromodulation studies. To characterize the acoustic pressure field at the source surface, we performed planar scans with a 0.2 mm needle hydrophone (Precision Acoustics, UK) in a water tank filled with distilled water. The hydrophone was positioned 18 mm from the transducer surface center (along the acoustic axis), and raster-scanned over a 66 mm × 66 mm plane with a 0.25 mm step length using a 3D scanning system (Velmex, US). Transducers were driven with a 0.5 ms sinusoidal burst waveform using a function generator (33500B, Keysight, US), and the electrical drive was set to 1 W using a power amplifier (1020L, E&I, US). Received signals were digitized with a PC oscilloscope (4224A, Picoscope, US) at 40 MHz, and band-pass filtered around the driven frequency.

The hologram was imported into the k-Wave Matlab toolbox, each hologram was back-propagated to a bowl detector (diameter = 60 mm, ROC = 75 mm), the center distance, grid spacing and time step were set to be the same as the experimental measurements, and the Courant–Friedrichs–Lewy (CFL) number was set at 0.15. The pressure distribution recorded on the bowl detector was taken as the transducer surface pressure field. The reconstructed bowl-surface field was then forward propagated to generate the free-field acoustic volume used as source validation, and as inputs for transcranial simulations (see Figure 2. a and b).

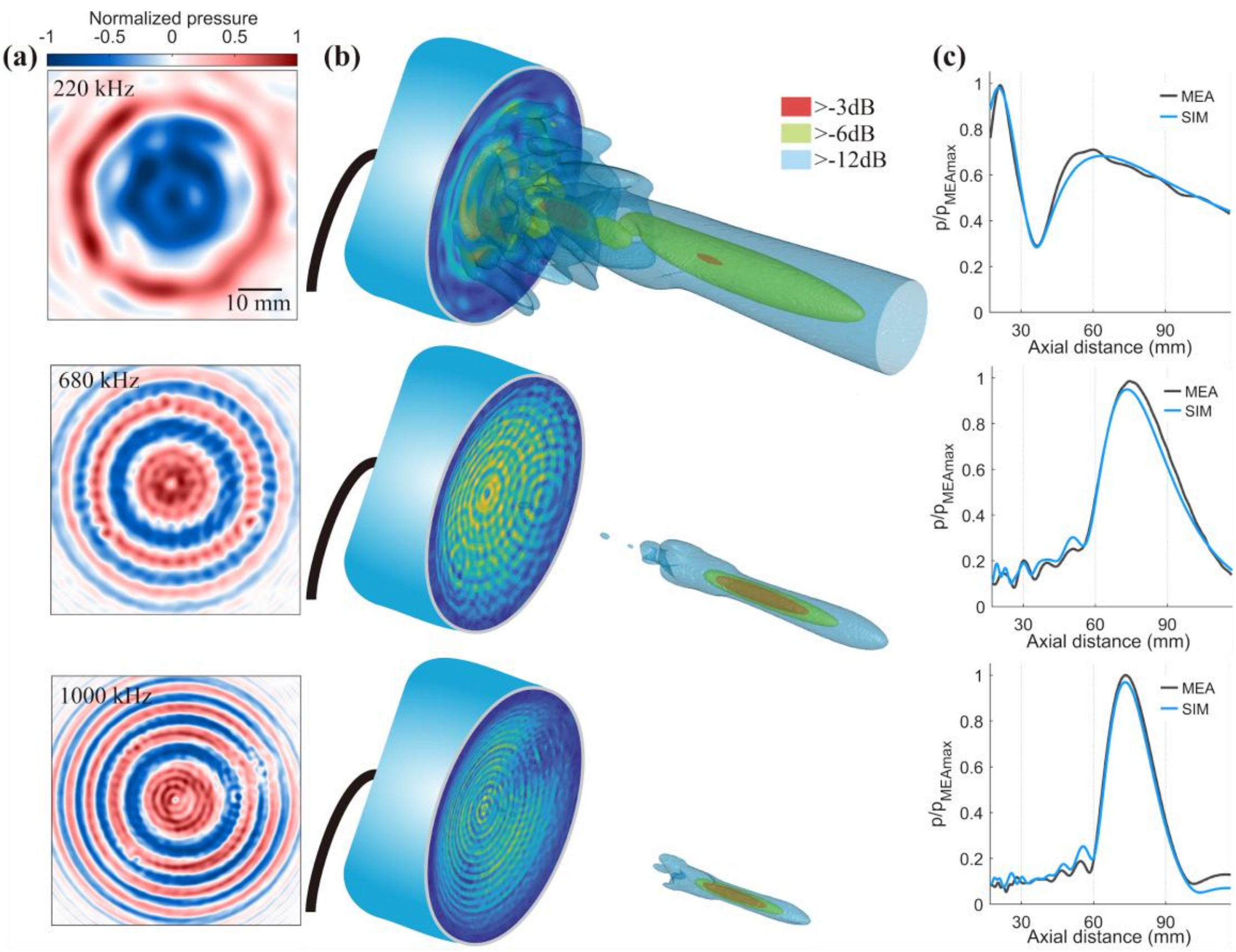

*Figure 2. (a) Acoustic holograms measured at transducer near-field. (b) Source surface acoustic pressure backward simulated from holograms, and their consequent forward simulated focal volume, thresholding at -3 dB, -6 dB and -12 dB. (c) Comparison of the axial pressure distribution between simulation and measurement, the pressure was normalized to the measured peak pressure.*

## Intracranial acoustic field characterization

To achieve rigid, repeatable alignment of each transducer to the target ROI on the skull, we designed a helmet-shaped transducer-skull scaffold for each ROI: The skull morphology was retrieved by thresholding using 3D Slicer, then the skull segments were processed using MeshLab to repair voids and reduce the vertices complexity. After which, the cleaned skull outer mesh was inverted to form the helmet inner surface. A clip mount was integrated to hold the transducer with designated distance from the housing to the ROI surface, and finally 3D printed. In addition, a helmet-transducer-water tank mounting frame was built to maintain the relative placement of the transducer in both free field and with-skull conditions (see Figure 3).

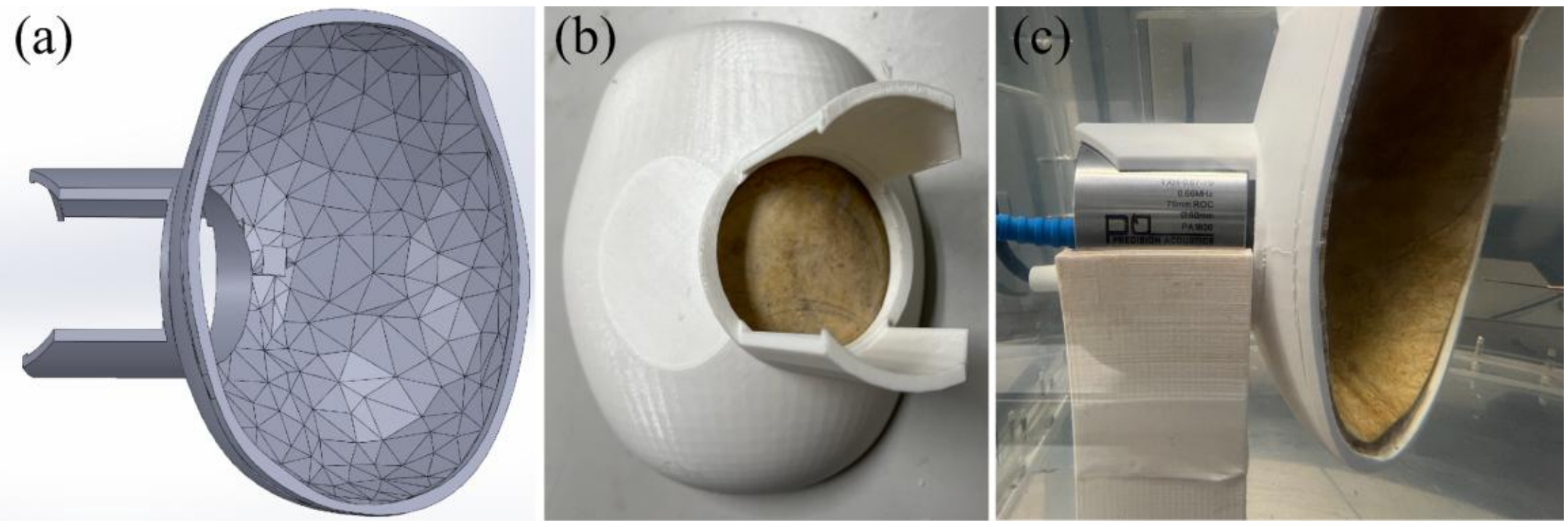

*Figure 3. A representative transducer-skull fixture design: (a): Scaffold mesh model tightly compatible with skull outer surface with transducer fixture design. (b): Integrate skull with the scaffold with an open window for ultrasound transmission through the ROI. (c): Underwater integration of transducer-skull scaffold, the scaffold was mounted on the scaffold holder fixed on the water tank. Hydrophone was scanning at the right side, intracranial space.*

Free-field mapping was performed at the far-field focal depths determined for 220 kHz, 680 kHz, 1000 kHz source: (63 mm, 75 mm, 74 mm), respectively. At each depth, we performed a raster scan over 30 mm × 30 mm for 220 kHz and 20 mm × 20 mm for 680 kHz and 1000 kHz, respectively. The same mapping geometry and depth were repeated with the integration of the skull. The resulting pressure maps were used for direct comparison with insertion effects.

To analyze focal behavior away from the measurement plane, each lateral complex pressure map was zero-padded to an effective 40 mm lateral extent, and then propagated using the k-Wave function angularSpectrumCW [33]. We performed bidirectional propagation by a total of 60 mm (from depth +45 mm to +105 mm) to estimate the focal-volume variation and to locate the 3D peak pressure within the evaluation domain.

For numerical prediction of the intracranial field, the measured source surface pressure was used as source input, each skull ROI segment was up sampled to 0.25 mm isotropic grid spacing and parameterized using the above described five skull-modelling techniques. The ROI was placed 10 mm in front of the source. The simulation domain comprised 512 × 288 × 288 voxels, covering approximately lateral -30 mm to +30 mm, and axial 0 mm to 120 mm.

## Accuracy evaluation metrics

Simulation (sim) and measurement (mea) were compared using four complementary metrics. Peak pressure loss ($p$) describes how much the maximum pressure is reduced after transmission through the skull. Intensity loss ($I$) estimates the corresponding reduction in acoustic energy delivery over the focal region. Focal volume ($V$) describes the size of the region enclosed by a chosen pressure threshold and therefore reflects predicted target coverage. Peak displacement ($D$) describes the movement of the intracranial ($IC$) pressure maximum relative to the free-field ($free$) focus and therefore reflects the targeting accuracy. The corresponding equations to evaluate errors are shown below:

**1.** Insertion loss errors: $\varepsilon_p = \frac{|p_{sim} - p_{mea}|}{p_{mea}} * 100\%$, $\varepsilon_I = \frac{|I_{sim} - I_{mea}|}{I_{mea}} * 100\%$,

where $p$ and $I$ are insertion-normalised pressure: $p = \frac{p_{IC}}{p_{free}}$ and intensity: $I = \frac{I_{IC}}{I_{free}}$.

Intensity on the mapping plane: $I = \frac{1}{A_{-6dB}} \int_{A_{-6dB}} \frac{p^2}{\rho c} dA$, where $A_{-6dB}$ is the -6 dB coverage of the local peak.

**2.** Focal volume variation: $\varepsilon_{V,\tau} = \frac{|(V_\tau)_{sim} - (V_\tau)_{mea}|}{(V_\tau)_{mea}} * 100\%$, $\tau \in \{-3, -6, -12\ dB\}$, where $V\tau$ is insertion-normalized volume: $V\tau = \frac{(V_\tau)_{IC}}{(V_\tau)_{free}}$.

**3.** The peak location displacement $\varepsilon_D = |(\Delta D)_{sim} - (\Delta D)_{mea}|$,

where $D$ is the vector distance from the free-field peak to the intracranial peak: $D = x_{IC} - x_{freefield}$ , and $\Delta D$ is the signed axial-lateral displacement: $\Delta D = sgn(D_{axial})\sqrt{{D_{axial}}^2 + {D_{lateral}}^2}$.

# Results

## Source validation in free-field

Validation of the numerically reconstructed sources was undertaken by comparing forward simulations with free-field hydrophone measurements acquired along the acoustic axis between 18 mm and 118 mm from each transducer in 0.5 mm increments. As shown in Figure 2.c, across the three frequencies, the simulated axial peak pressure was approximately 3% to 7% lower than experimental values. As shown in Table 1, three-dimensional pressure distributions derived from holographic reconstruction of focal field pressure mapping provided a benchmark for the simulated focal field. The volumetric focal errors remained around or below 10% for the -3 dB, -6 dB and -12 dB isosurfaces, and the residual discrepancy in the position of the spatial peak pressure limited at approximately 2 to 3 millimeters. These small deviations indicate that the reconstructed source fields adequately reproduce the essential free-field behavior of the transducers with sufficient accuracy to justify their use in the subsequent transcranial simulations. Nevertheless, these errors were eliminated when comparing the intracranial field by normalizing to its free-field values, to show only the acoustic field variation caused by the presence of the skull.

*Table 1. Signed (sgn) errors of simulation in free field compared to focal field mapping (p and I) and holographic reconstruction (V and D)*

| Free field (sgn) | 220 kHz | 680 kHz | 1000 kHz |
|---|---|---|---|
| $\varepsilon_p$ (%) | -2.8 | -6.6 | -7.4 |
| $\varepsilon_I$ (%) | -8.3 | -11.3 | -14.0 |
| $\varepsilon_{V,-3dB}$ (%) | -2.5 | 3.5 | 2.7 |
| $\varepsilon_{V,-6dB}$ (%) | 10.3 | 6.1 | -6.0 |

| $\varepsilon_{V,-12dB}$ (%) | 2.0 | 2.3 | -9.2 |
|---|---|---|---|
| $\varepsilon_{D}$ (mm) | 3.2 | -2.6 | -1.8 |

## Representative intracranial field variability

A representative example is shown in Figure 4 for a RoI located in the frontal bone of specimen S2 and modelled using the 3L-D strategy. Distortion was evident at all three frequencies. In each case, the intracranial focus was shallower than the free-field focus (indicated by black dot), and significant wave interactions occurred both in the transducer-skull spacing and within the skull thickness. At 220 kHz, the pressure in the gap region exceeded the intracranial focal pressure, indicating strong low-frequency superposition. At 680 kHz and 1000 kHz, clear sidelobes appeared, suggesting beam splitting associated with the frontal crest.

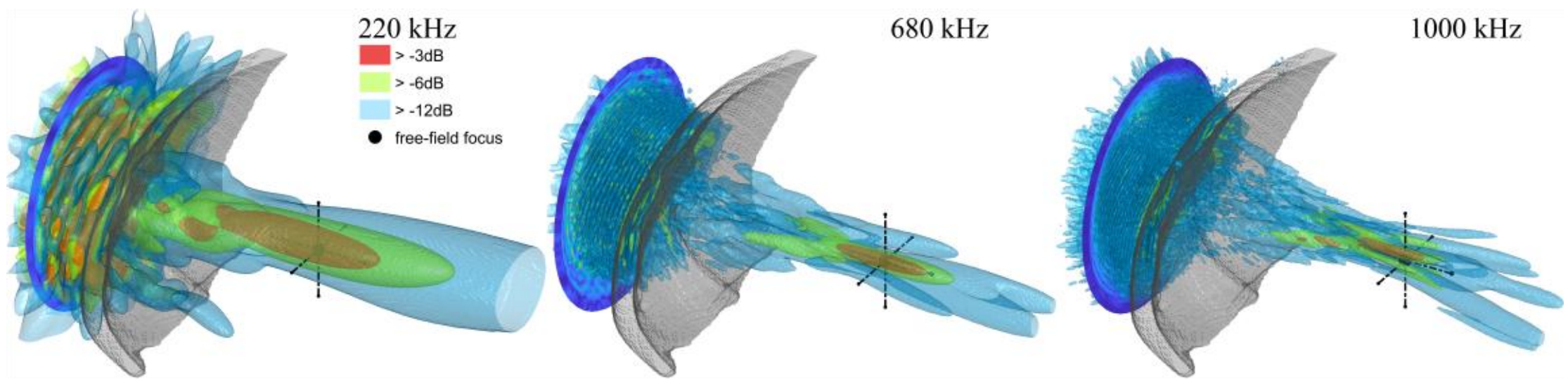

*Figure 4. Simulated focal field with a piece of RoI at S2 frontal bone at 220 kHz, 680 kHz and 1000 kHz. Skull was modelled using 3L-D strategy. The pressure levels are normalized to peak intracranial field, black dot represent the focus at free-field.*

Figure 5 compares the five skull-modelling strategies for the same RoI. Although the models produce markedly different distributions of properties, their intracranial focal-pressure fields remained broadly similar. More substantial divergence emerges at 680 kHz and 1000 kHz, where the 3L-F, 3L-D, and 1L-F models reproduce enlarged sub-focal lobes, whereas the two linear-mapping models yield more compact foci. Quantitatively, no single model achieves uniform superiority across all three frequencies, the relative peak pressure errors extended between +1.73% (3L-F) to -21.5% (1L-F), +13.8% (LM-Da) to -10.0% (3L-F), and +32.5% (3L-D) to -4.8% (1L-F) at the three frequencies, respectively.

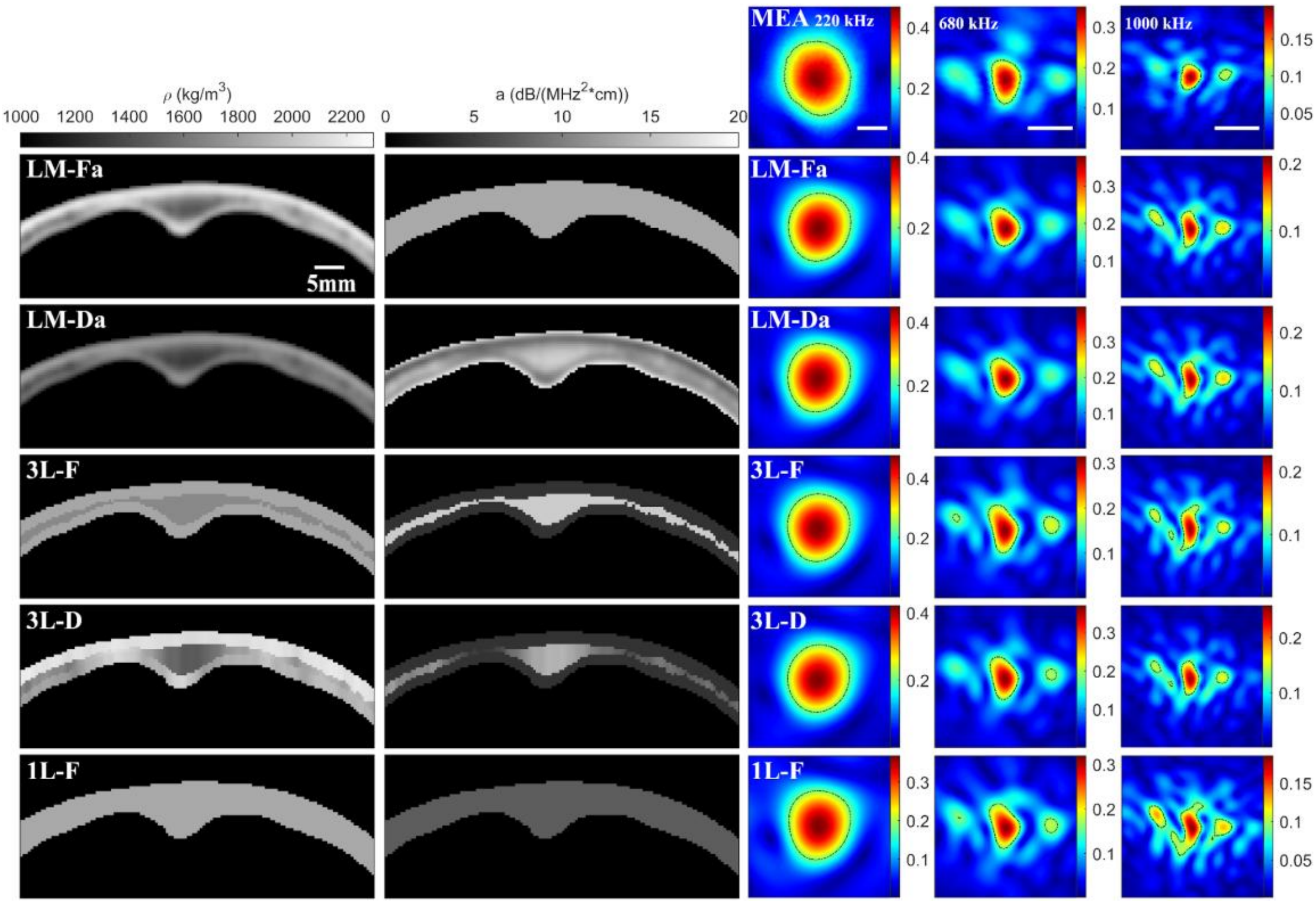

*Figure 5 Output of five models from a representative ROI at S2 frontal bone and its corresponding simulated cross-sectional focal field at 220 kHz, 680 kHz and 1000 kHz, compared with respective measured fields (MEA). Dashed contours indicate regions above the -6 dB pressure threshold.*

## Grouped insertion loss and focal metrics

The 19 ROIs selected for insertion experiments sampled a wide anatomical space: skull thickness ranged from 3.1 mm (S1) to 11.2 mm (S5) (grouped mean±std = 6.3±2.3 mm); bulk HU density from 781 HU (S1) to 1445 HU (S5) (grouped = 1092±213 HU); and bulk SDR values between 0.39 (S6) to 0.73 (S1) (grouped = 0.58±0.08). Grouped median results are summarized in Figure 6 to Figure 9 and in Table 2.

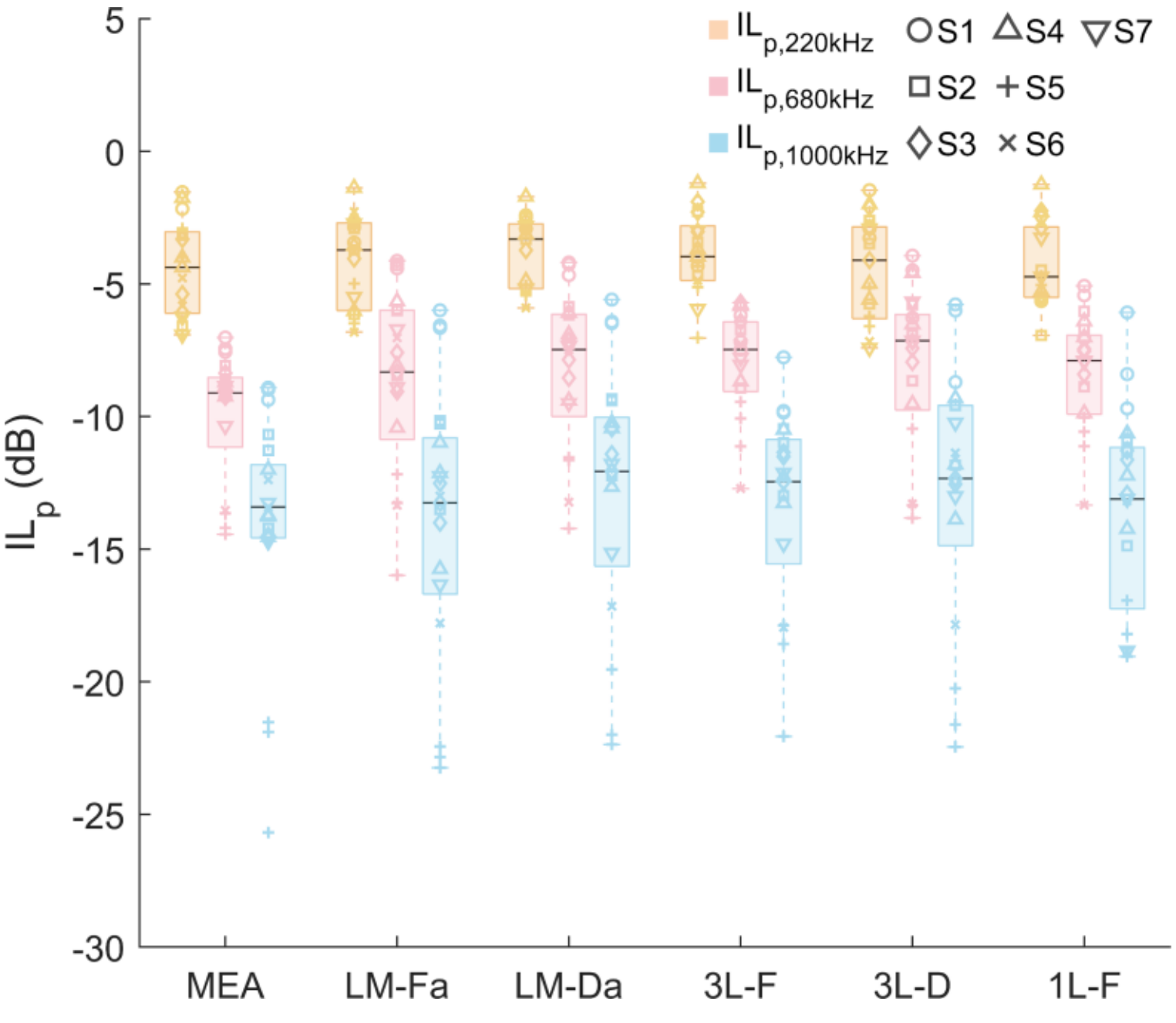

*Figure 6 Measured (MEA) and simulated intracranial peak pressure loss across 19 ROIs from seven skull specimens*

*at 220 kHz, 680 kHz and 1000 kHz frequencies. The results are shown in median values and 25% and 75% interquartile range (IQR), with whiskers extending to beyond the 1.5 IQR.*

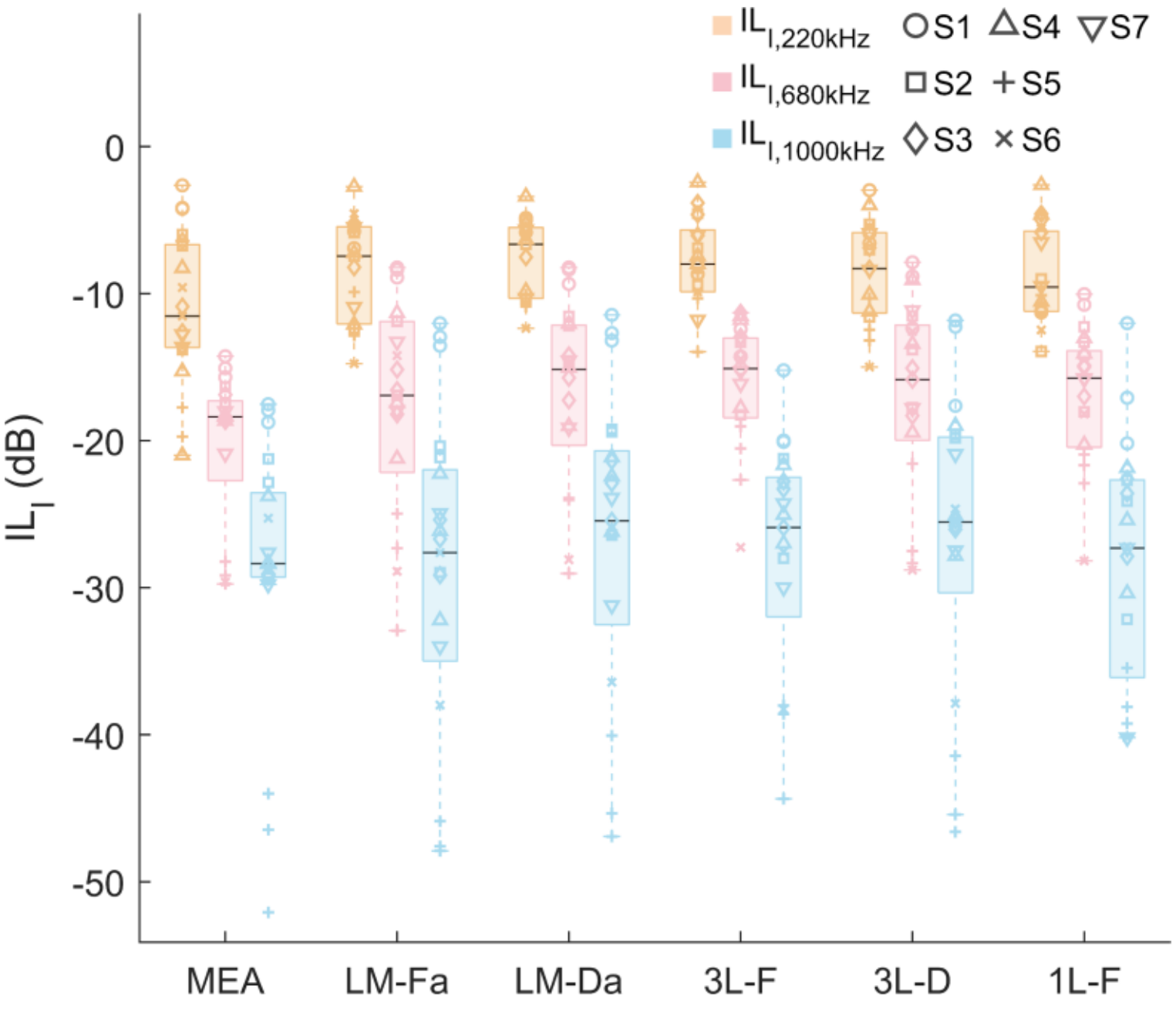

*Figure 7 Measured (MEA) and simulated intracranial intensity (over -6 dB cross-sectional beam area) loss across 19 ROIs from seven skull specimens at 220 kHz, 680 kHz and 1000 kHz frequencies.*

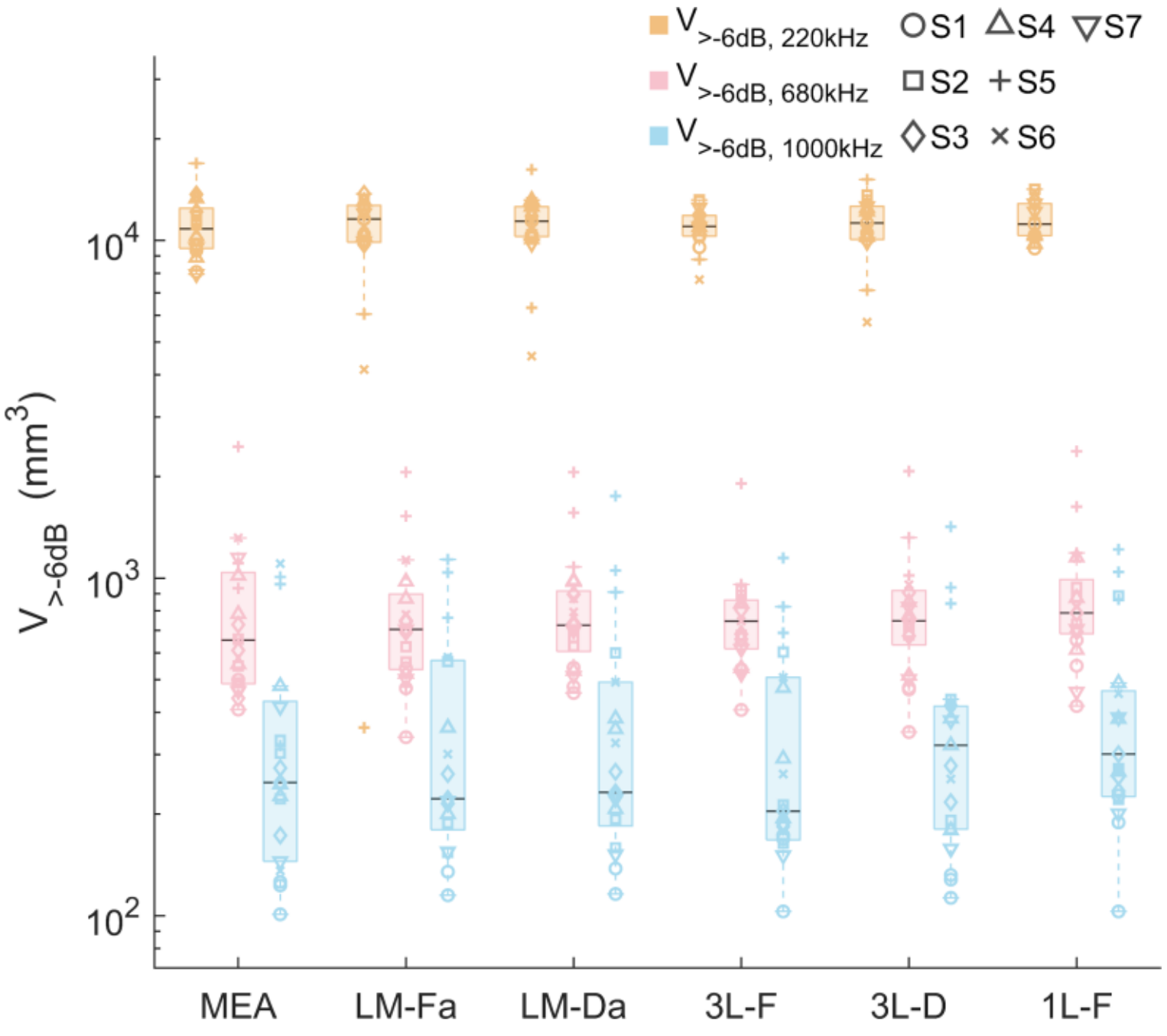

*Figure 8 Hologram propagated from 2D measurement and simulated intracranial acoustic volumetric coverage (over -6 dB) across 19 ROIs from seven skull specimens at 220 kHz, 680 kHz and 1000 kHz frequencies.*

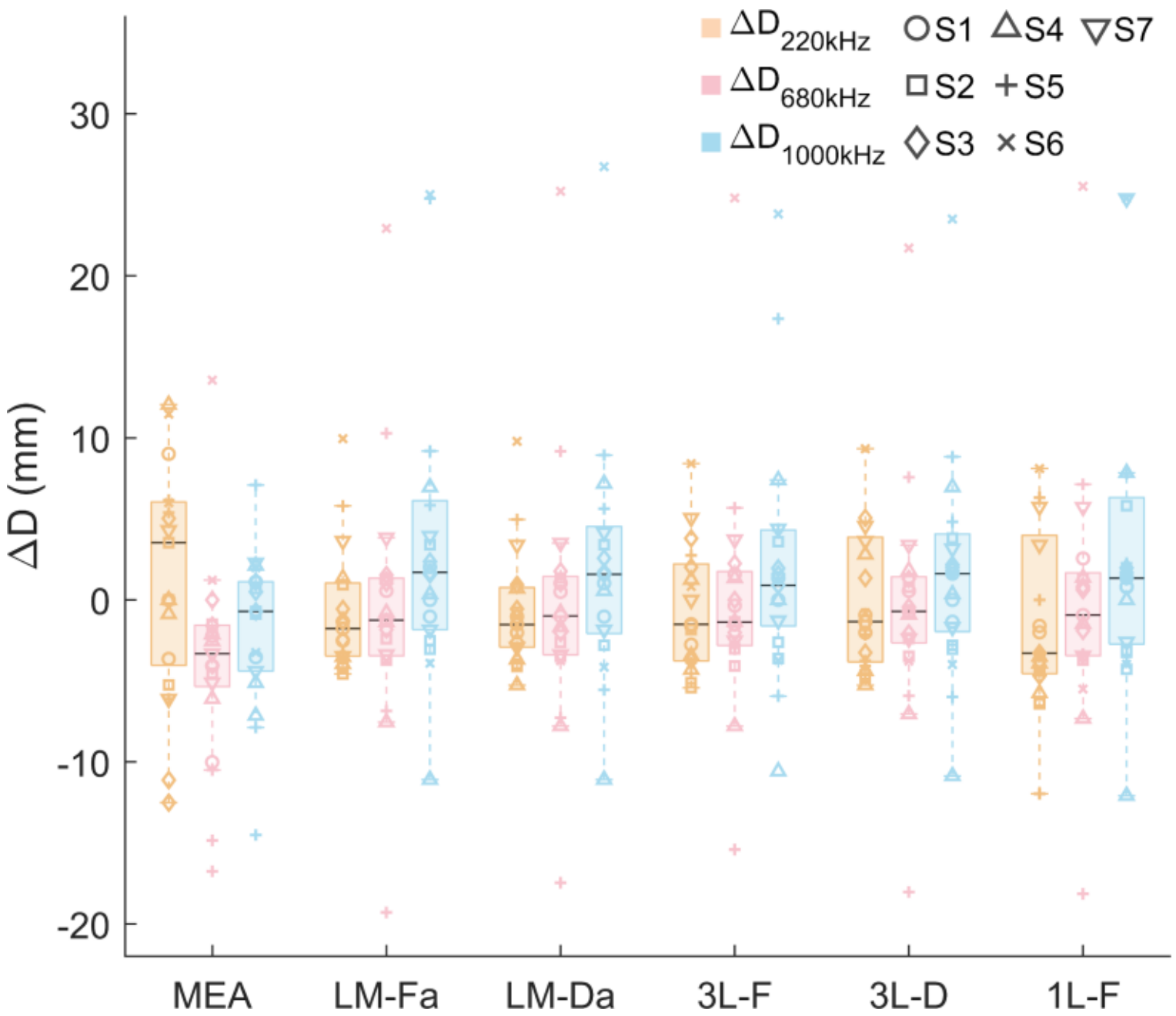

*Figure 9 Hologram propagated from 2D measurement, and simulated intracranial peak pressure distance deviation to their free-field focus across 19 ROIs from seven skull specimens at 220 kHz, 680 kHz and 1000 kHz frequencies.*

Compared with measurement, the simulated normalized peak-pressure loss (Figure 6) at the free-field focal depth differed by median values from -0.3 to +2.0 dB, across the three frequencies, the corresponding intensity loss (Figure 7) difference ranged from +0.8 to +4.9 dB. These systematically smaller simulated losses indicate that the current skull modelling strategies tended, on average, to underestimate attenuation or overestimate transmitted energy. This bias is more consequential for intensity than for pressure, because intensity depends on the square of pressure and is further modulated by beam area.

The absolute focal volume discrepancy (Figure 8) above the -6 dB threshold decreased with increasing frequency: at 220 kHz the median discrepancy remained on the order of several hundred cubic millimeters, at 680 kHz it fell to the low hundreds, and at 1000 kHz it was typically only a few dozen cubic millimeters. When expressed relative to the measured focal volumes, the maximum discrepancy rose from +6.9% at 220 kHz to +29.0% at 1000 kHz. The direction of this error was frequency dependent. At 220 and 680 kHz, all five simulations produced larger median -6 dB focal volumes than measured, indicating overestimation of target coverage. At 1000 kHz, the direction became model dependent, with some models underestimating and others overestimating the measured focal volume. Thus, although absolute focal-volume error diminished with frequency, the relative uncertainty remained substantial because the measured focal volume was much smaller at higher frequency.

The simulated intracranial pressure maxima are on average located superficial to free-field peaks at 220 kHz and 680 kHz, whereas the relationship reversed at 1000 kHz (Figure 9). In the measured data, the median displacement of the intracranial peak relative to the free-field focus was +3.5 mm at 220 kHz, -3.3 mm at 680 kHz and -0.7 mm at 1000 kHz. The simulations predicted median displacements ranging from -3.3 to -1.3 mm, -1.4 to -0.7 mm and +0.9 to +1.7 mm at the three frequencies, respectively.

Therefore, the typical disagreement in focal position was on the order of several millimeters. This is small relative to the overall propagation distance, but it remains relevant when interpreting target coverage in small or anatomically adjacent brain regions.

*Table 2 Measured (MEA) and simulated intracranial acoustic field results at three frequencies, report as median and IQR (in []).*

| | MEA | LM-Fa | LM-Da | 3L-F | 3L-D | 1L-F |
|---|---|---|---|---|---|---|
| 220 kHz | | | | | | |
| $IL_p$ (dB) | -4.4 [-6.1 -3.0] | -3.7 [-6.0 -2.7] | -3.3 [-5.2 -2.7] | -4.0 [-4.8 -2.8] | -4.1 [-6.3 -2.8] | -4.7 [-5.5 -2.8] |
| $IL_I$ (dB) | -11.5 [-13.7 -6.7] | -7.5 [-12.1 -5.5] | -6.6 [-10.3 -5.5] | -8.0 [-9.9 -5.7] | -8.3 [-11.3 -5.8] | -9.6 [-11.2 -5.7] |
| $V_{-6dB}$ (mm$^3$) | 10827 [9468 12460] | 11573 [9885 12701] | 11413 [10275 12588] | 11004 [10300 11852] | 11261 [10072 12619] | 11179 [10337 12860] |
| $\Delta D$ (mm) | 3.5 [-4.0 6.0] | -1.8 [-3.5 1.0] | -1.5 [-2.9 0.8] | -1.5 [-3.8 2.2] | -1.3 [-3.8 3.9] | -3.3 [-4.6 4.0] |
| 680 kHz | | | | | | |
| $IL_p$ (dB) | -9.1 [-11.2 -8.5] | -8.3 [-10.8 -6.0] | -7.5 [-10.0 -6.1] | -7.5 [-9.1 -6.4] | -7.1 [-9.8 -6.2] | -7.9 [-9.9 -6.9] |
| $IL_I$ (dB) | -18.4 [-22.7 -17.3] | -16.9 [-22.2 -11.9] | -15.2 [-20.3 -12.1] | -15.1 [-18.4 -13.0] | -15.9 [-20.0 -12.1] | -15.7 [-20.4 -13.9] |
| $V_{-6dB}$ (mm$^3$) | 655 [487 1040] | 705 [536 896] | 725 [608 916] | 746 [617 861] | 747 [634 919] | 788 [684 990] |
| $\Delta D$ (mm) | -3.3 [-5.3 -1.6] | -1.3 [-3.5 1.3] | -1.0 [-3.4 1.5] | -1.4 [-2.8 1.7] | -0.7 [-2.7 1.4] | -0.9 [-3.4 1.7] |
| 1000 kHz | | | | | | |
| $IL_p$ (dB) | -13.4 [-14.6 -11.8] | -13.3 [-16.7 -10.8] | -12.1 [-15.6 -10.0] | -12.5 [-15.6 -10.9] | -12.3 [-14.9 -9.6] | -13.1 [-17.2 -11.2] |
| $IL_I$ (dB) | -28.4 [-29.3 -23.5] | -27.6 [-35.0 -22.0] | -25.5 [-32.5 -20.7] | -25.9 [-32.0 -22.5] | -25.5 [-30.3 -19.8] | -27.3 [-36.1 -22.7] |
| $V_{-6dB}$ (mm$^3$) | 248 [145 431] | 222 [180 570] | 232 [185 492] | 204 [168 508] | 320 [181 418] | 301 [226 464] |
| $\Delta D$ (mm) | -0.7 [-4.4 1.1] | 1.7 [-1.8 6.1] | 1.6 [-2.1 4.5] | 0.9 [-1.6 4.3] | 1.6 [-2.0 4.1] | 1.3 [-2.7 6.3] |

## Model-specific error analysis

Table 3 summarizes the averaged absolute errors obtained when the five skull modelling approaches are benchmarked against the 19 experimental intracranial datasets at 220 kHz, 680 kHz and 1000 kHz. Across all frequencies and models, peak pressure errors cluster in the range of 20% to 31%, with frequency-averaged values of 23±16% at 220 kHz, 25±18% at 680 kHz, and 27±25% at 1000 kHz. The corresponding intensity errors are in the range of 41% to 77%, with frequency-averaged values of 60±51%, 63±52%, and 60±76%, respectively.

The model ranking was not identical at every frequency. For peak pressure, the lowest mean error was obtained by 3L-D at 220 kHz, by 1L-F at 680 kHz and by LM-Fa at 1000 kHz. For the intensity-related metric, the lowest mean error was obtained by 3L-D at 220 kHz, by 1L-F at 680 kHz and by LM-Fa at 1000 kHz. Nevertheless, LM-Fa was consistently below the ensemble mean for both peak-pressure and intensity errors, and gave the lowest frequency-averaged errors across the full dataset. Conversely, 1L-F showed the largest errors in several comparisons, particularly at 220 kHz and 1000 kHz, although it was not uniformly the poorest model at every frequency or for every metric.

Volumetric accuracy was assessed at -3 dB, -6 dB and -12 dB pressure thresholds. At 220 kHz, the model-averaged errors were 33±39%, 21±17% and 15±10%, respectively. At 680 kHz, these values were 27±22%, 28±21% and 16±14%, respectively. At 1000 kHz, they increased to 44±51%, 50±53% and 29±32%, respectively. The 1L-F model produced the largest error in most volumetric comparisons, especially at -3 dB and -12 dB; however, exceptions were present, and the remaining four approaches were broadly comparable.

Spatial fidelity, quantified by the signed axial-lateral displacement, showed only modest differences among models. Mean displacement errors were 6.6±4.3 mm at 220 kHz, 5.2±4.7 mm at 680 kHz, and 5.6±7.3 mm at 1000 kHz. LM-Da gave the smallest frequency-averaged displacement, but the advantage was less than 1 mm and therefore not practically decisive.

*Table 3 Average intracranial errors of simulated results using 5 skull modelling methods compared to the measurement at 220 kHz, 680 kHz and 1000 kHz.*

| 220 kHz | LM-Fa | LM-Da | 3L-F | 3L-D | 1L-F |
|---|---|---|---|---|---|
| $\varepsilon_p$ (%) | 22±17 | 21±16 | 25±19 | 20±16 | 28±14 |
| $\varepsilon_I$ (%) | 54±47 | 57±53 | 65±58 | 50±47 | 72±51 |
| $\varepsilon_{V,-3dB}$ (%) | 32±32 | 31±33 | 31±40 | 32±31 | 37±59 |
| $\varepsilon_{V,-6dB}$ (%) | 25±25 | 22±18 | 11±22 | 22±13 | 20±14 |
| $\varepsilon_{V,-12dB}$ (%) | 15±10 | 15±10 | 13±8 | 14±8 | 16±11 |
| $\varepsilon_D$ (mm) | 6.4±3.9 | 6.3±4.2 | 7.1±4.6 | 6.7±4.8 | 6.3±4.3 |
| 680 kHz | | | | | |
| $\varepsilon_p$ (%) | 24±17 | 25±14 | 27±21 | 27±22 | 23±17 |
| $\varepsilon_I$ (%) | 59±46 | 61±41 | 72±62 | 65±67 | 57±46 |
| $\varepsilon_{V,-3dB}$ (%) | 26±20 | 25±22 | 27±22 | 25±20 | 34±27 |
| $\varepsilon_{V,-6dB}$ (%) | 25±20 | 25±20 | 27±21 | 25±22 | 36±24 |
| $\varepsilon_{V,-12dB}$ (%) | 16±18 | 14±13 | 13±10 | 17±15 | 19±11 |
| $\varepsilon_D$ (mm) | 5.2±4.9 | 5.1±4.8 | 5.0±4.7 | 5.1±4.4 | 5.6±4.9 |
| 1000 kHz | | | | | |
| $\varepsilon_p$ (%) | 22±12 | 31±16 | 26±33 | 27±19 | 30±39 |
| $\varepsilon_I$ (%) | 41±28 | 67±38 | 62±92 | 56±46 | 77±129 |
| $\varepsilon_{V,-3dB}$ (%) | 40±43 | 42±50 | 43±53 | 40±43 | 56±65 |
| $\varepsilon_{V,-6dB}$ (%) | 43±38 | 51±48 | 42±41 | 49±46 | 67±83 |
| $\varepsilon_{V,-12dB}$ (%) | 25±24 | 27±25 | 28±28 | 25±30 | 39±47 |
| $\varepsilon_D$ (mm) | 6.2±9.6 | 4.8±6.0 | 6.0±8.1 | 4.6±5.3 | 6.3±7.2 |

The two skull specimen cohorts differ markedly in skull properties. Historical calvaria ROIs had mean thickness of 5.8±1.7 mm, mean CT HU value of 987±139 HU, and SDR of 0.54±0.09. whereas Thiel-embalmed ROIs were thicker and denser, with values of 8.5± 2.5 mm, 1359±88 HU, and 0.62±0.07 respectively. These differences were reflected in the larger energy-loss errors observed in the Thiel group, which is summarized in Table 4.

*Table 4. Pressure and intensity errors stratified by specimen condition.*

| | Historical ROIs | | | Thiel ROIs | | |
|---|---|---|---|---|---|---|
| | 220 kHz | 680 kHz | 1000 kHz | 220 kHz | 680 kHz | 1000 kHz |
| $\varepsilon_p$ (%) | 23±18 | 22±16 | 25±16 | 24±11 | 31±22 | 32±38 |
| $\varepsilon_I$ (%) | 52±46 | 54±44 | 51±40 | 77±57 | 81±64 | 81±121 |

# Discussion

The present work examined five skull-modelling strategies implemented in k-Wave and benchmarked their intracranial field predictions against hydrophone measurements. A key finding is that all five skull models reproduced the broad intracranial beam pattern, but none provided consistently accurate quantitative predictions across pressure, intensity, focal volume and focal displacement. Mean peak pressure errors remained on the order of 20%-31%, but intensity errors were substantially larger because they were amplified by the nonlinear dependence on pressure and by variation in beam width. The simulations generally predicted less insertion loss than measured, suggesting that skull attenuation, soft-tissue effects or CT-to-acoustic property mappings remain incompletely represented. If a similar bias occurs in simulation-guided human tFUS studies, selected driving levels may under-deliver the intended intracranial pressure or intensity, potentially reducing the likelihood of observing neuromodulatory effects. Beyond this exposure uncertainty, volumetric discrepancies at the -6 dB threshold were also substantial, especially at higher frequency when the focal volume became smaller and therefore more sensitive to modest modelling error.

## Limitations of skull modelling strategies

In principle, the fully voxelized LM-Da approach might be expected to perform best, because it retains local heterogeneity in density, sound speed, and attenuation. However, the present data does not support that expectation. When an alternative attenuation parameter set $a_{skull_max}$=8.7 dB/(cm·MHz$^{1.43}$), $a_{skull_min}$=4 dB/(cm·MHz$^{1.43}$) [26] was tested at 680 kHz, the median deviations in peak pressure and intensity increased from +1.6 dB to +1.9 dB, and from +3.2 dB to +4.0 dB, respectively. This suggests that uncertainty in attenuation assignment may outweigh the theoretical benefit of retaining fully voxel-wise heterogeneity. Notwithstanding, the present comparison did not isolate attenuation independently from the associated sound speed and density transforms, so that conclusion remains provisional.

Blurred HU gradients at the air-bone boundary are also likely to render the outer and inner cortical tables artificially low in density, thereby exaggerating attenuation in the linear-mapping models. Reassignment of these transitional voxels to cortical-like properties may therefore improve predictive performance. This interpretation is supported by the good performance of LM-Fa. Despite its simplicity, LM-Fa produced lower average pressure error across the three frequencies. Conversely, the homogeneous 1L-F model was generally the least accurate, indicating that some anatomical stratification is necessary even if full voxel-wise heterogeneity does not yet yield a clear advantage under the present parameter mappings. The difference between 3L-F and 3L-D further suggests that model structure alone is not sufficient; the accuracy of the mapped acoustic properties is at least as important as the geometric fidelity of the skull representation.

Comparison with a benchmark 3L-F configuration also supports this conclusion [25]. The benchmark setup used attenuation coefficients equivalent to 16 and 32 dB/(cm·MHz$^2$) for cortical and trabecular bone, respectively. In our dataset, this configuration produced peak pressure levels lower than measurement by -0.9 dB ($\varepsilon_p$=23±19%), -1.6 dB (20±15%), and -5.2 dB (46±19%) at 220 kHz, 680 kHz and 1000 kHz, respectively. Although this stronger attenuation reduced error at lower frequencies, it produced substantially larger error at 1000 kHz. If such a model were used to select transducer input levels, it could therefore potentially lead to overcompensation of the input pressure at higher frequency.

As shown in Table 4, Thiel-embalmed ROIs have on average a greater energy loss error. This may stem from specimen physical differences: as the Thiel specimens have greater thickness and density, but the present data cannot separate the effects of morphology, preservation state and unmodelled soft-tissue layers. Currently, direct acoustic property data for Thiel-embalmed skulls are limited. For comparison, formalin-fixed human skulls have reported longitudinal attenuation values of approximately 8.83 dB/(cm·MHz$^{1.43}$) [26,34,35]. Also, the embalmed material retains attached scalp and dura mater, media that were omitted from the simulations yet participate in the experimental transmission path. Their combined absorption and impedance mismatch would elevate the measured insertion loss relative to a bone-only model. Taken together, these observations indicate that bone-only simulations may be insufficient for quantitative neuromodulation planning whenever soft-tissue layers contribute materially to the acoustic path. Incorporating specimen-specific skin and meningeal properties, either by explicit segmentation or by empirical correction factors, should therefore form part of future high-fidelity modelling efforts.

Volumetric and spatial predictions add further caution. Errors at the -3 dB and -6 dB thresholds were substantial, particularly at 1000 kHz, indicating uncertainty in the predicted extent of effective stimulation. Overestimation risks exposing unintended tissue, whereas underestimation risks failing to cover the intended target. Mean axial-lateral displacement errors were smaller and similar across methods, but individual outliers can still be important for patient-specific planning, especially in thick frontal bone or highly curved temporal regions. These uncertainties were also exacerbated by experimental constraints: the lateral mapping window of approximately 20-30 mm spanned only a little more than twice the focal width, limiting capture of the full field. In future work, larger mapping windows and hologram acquisition at a slightly defocused plane may preserve more peripheral phase information, although this must be balanced against acquisition time.

## Systematic uncertainties

Source characterization was based on backward propagation of planar acoustic holograms to reconstruct the bowl-surface pressure distribution. Although this approach preserves empirical phase information, it can still introduce systematic underestimation of radiated energy and small field distortions. First, the planar scan assumes that the incident wavefront is sampled under approximately normal incidence, whereas the hydrophone tip encounters a distribution of incidence angles; at 680 kHz, a 20° tilt reduced the measured peak by approximately 3.5%. Second, residual reflections from the sensing tip, even after angular normalization and optimization, contributed to near-field errors of approximately 1.3%. Third, peripheral wave information can be lost when the recorded hologram is back-propagated to a detector aperture that is marginally smaller than the effective radiating surface. Additional losses may arise from hydrophone spatial averaging and finite-bandwidth filtering during signal processing.

Alternative simulation approaches often bypass explicit source reconstruction by projecting the measured hologram to a prescribed plane and treating it as a mass source (calculateMassSourceCW) [33], or computing an idealized complex pressure distribution directly from the nominal transducer geometry (kWaveArray) [13]. These approaches are computationally convenient, but they also do not explicitly represent the finite acoustic impedance of the piezoceramic bowl or its interaction with waves reflected from the skull surface. This omission may be acceptable when the skull is remote from the transducer or when reflected energy is negligible, but it becomes more questionable for closely spaced, multi-cycle sonication in the sub-MHz regime, where standing-wave and superposition effects can be pronounced [15].

Representing the piezoceramic layer explicitly at the bowl source [29,36] may therefore provide a more physically faithful boundary condition. However, this also introduces an additional interface-loss term because waves generated within the piezoceramic must transmit into water before reaching the skull. The magnitude of this loss is not yet well constrained. A more rigorous future implementation could combine a measured bowl-surface pressure map with an explicit piezoceramic layer, while treating the piezoceramic-water transmission loss as a calibrated correction rather than as an uncontrolled modelling artefact. A further, albeit smaller, source of uncertainty stems from rotational alignment: the bowl-surface maps in Figure 2. a-b show intrinsic asymmetry, so the simulated field will not exactly match the experiment unless the transducer is immobilized about its acoustic axis during both source characterization and skull transmission measurements.

## Conclusion

This study provides a systematic evaluation of five skull-modelling schemes for k-Wave simulations of tFUS. Using 19 ROIs from five historical and two Thiel-embalmed human skulls, we compared simulated intracranial pressure fields with hydrophone measurement at 220 kHz, 680 kHz and 1000 kHz. After source characterization by acoustic holography, model performance was quantified in terms of pressure error, intensity error, focal volume error and spatial peak displacement.

Across all frequencies, mean peak pressure errors ranged from 20% to 31%, whereas intensity errors were larger, ranging from 41% to 77%. Volumetric errors were strongly threshold and frequency dependent, with the -6 dB error increasing from approximately 11% at 220 kHz to 67% at 1000 kHz. LM-Fa gave the lowest frequency-averaged pressure and intensity errors and remained below the ensemble mean at each frequency, but it was not the best-performing method in every individual comparison. No model showed a decisive advantage across all performance dimensions. Thiel-embalmed specimens showed larger energy-loss errors than historical calvaria, particularly for intensity and at higher frequencies, likely reflecting their greater thickness, possible impedance mismatch differences and the contribution of unmodelled soft-tissue layers.

Overall, skull parameterization remains a major source of uncertainty in quantitative tFUS simulation. Increasing anatomical detail alone is insufficient unless CT-derived density, sound speed and attenuation assignments are physically appropriate for the specimen and frequency range. The five evaluated approaches captured the gross intracranial beam pattern, but none provided uniformly reliable quantitative accuracy for dose-controlled neuromodulation without further validation or correction. For current simulation-guided tFUS studies, the observed tendency to underestimate insertion loss suggests that intended intracranial exposure may be under-delivered in some cases, potentially contributing to variable neuromodulatory outcomes. Future work should improve attenuation laws, interface modelling and patient-specific verification.

# Acknowledgements

The experimental work was conducted in University of Dundee, and the author expresses deep appreciation to Mr. Tyler Halliwell, Dr. Tom Gilbertson, Dr. Isla Barnard, Professor. Andreas Melzer, and Mr. Alan Webster for technical and academic support.

# Author declarations

## Conflict of Interests

The authors have no conflicts of Interest to disclose.

## Ethics approval

Approval of the use of human tissue specimens in the study was given by the University of Dundee Thiel Advisory group. This research complies with the Anatomy Act (1984) and the Human Tissue (Scotland) Act 2006.

# Data availability

The raw skull data is not publicly available due to ethical restrictions